\documentclass[journal]{IEEEtran}
\usepackage{graphicx}
\usepackage{xcolor}
\usepackage{subcaption}
\usepackage{amsmath}
\usepackage{multirow}

\usepackage{todonotes}
\usepackage{url}

  \usepackage[nocompress]{cite}

\begin{document}
%
\title{Acoustic and linguistic representations for speech continuous emotion recognition in call center conversations.}

\author{Manon Macary, Marie Tahon, Yannick Estève, Daniel Luzzati
\thanks{M. Macary, M. Tahon and D. Luzzati are with the LIUM, Le Mans Université, France.}
\thanks{Y. Estève is with LIA, Avignon Université, France.}
\thanks{Manon Macary is also employed with Allomedia, Paris, France.}}

\markboth{.}%
{Macary \MakeLowercase{\textit{et al.}}: Mutual impact of acoustic and linguistic representations for continuous emotion recognition in call center conversations.}

\IEEEtitleabstractindextext{%
\begin{abstract}
The goal of our research is to automatically retrieve the satisfaction and the frustration in real-life call-center conversations. 
This study focuses an industrial application in which the customer satisfaction is continuously tracked down to improve customer services.
To compensate the lack of large annotated emotional databases, we explore the use of pre-trained speech representations as a form of transfer learning towards AlloSat corpus.
%
Moreover, several studies have pointed out that emotion can be detected not only in speech but also in facial trait, in biological response or in textual information. 
In the context of telephone conversations, 
we can break down the audio information into acoustic and linguistic by using the speech signal and its transcription. 
Our experiments confirms the large gain in performance obtained with the use of pre-trained features.
Surprisingly, we found that the linguistic content is clearly the major contributor for the prediction of satisfaction and best generalizes to unseen data.
Our experiments conclude to the definitive advantage of using CamemBERT representations, however the benefit of the fusion of acoustic and linguistic modalities is not as obvious.
With models learnt on individual annotations, we found that fusion approaches are more robust to the subjectivity of the annotation task.
This study also tackles the problem of performances variability and intends to estimate this variability from different views: weights initialization, confidence intervals and annotation subjectivity.
A deep analysis on the linguistic content investigates interpretable factors able to explain the high contribution of the linguistic modality for this task.
\end{abstract}

}

\maketitle

\IEEEdisplaynontitleabstractindextext
\IEEEpeerreviewmaketitle

\section{Introduction}
\label{sec:introduction}

\IEEEPARstart{N}{owadays}, relations between customers and companies are increasingly based on call centers~\cite{Cheong2008}. 
Within these structures, massive speech data is collected and automatically processed everyday by companies, since such data contains crucial information for these companies to improve their commercial relations with customers.
With the huge improvements in Automatic Speech Recognition and Spoken Language Understanding processing, it is now possible to extract automatically linguistic and semantic information for speech analytics.
In addition, paralinguistic cues can be useful to evaluate the customer level of commitment or attention to the agent discourse. 
One of the main paralinguistic cue of interest in such speech data is the emotional state of the speaker. 
In particular, frustration and satisfaction hold key factors of the customer relationship, and more precisely their evolution according time during the conversation.
In this paper, we focus on the automatic continuous extraction of such factors in the whole speech conversation.


Emotional states have been extensively studied and many theories exist~\cite{Russell1980, Scherer2005}. 
Among these, the continuous theory, also called dimensional theory, has been introduced by Wundt et al.~\cite{Wundt1902} and Scholsberg~\cite{Schlosberg1954}, and consider that all affective states arise from independent fundamental neurophysiological systems. 
According to this authors, these systems can be defined by three independent dimensions characterized by their extremum values: pleasant-unpleasant, tension-relaxation, and excitation-calm. 
These three dimensions were soon-to-be found overlapping.
Russell~\cite{Russell1980} introduced the circumplex model in which emotion categories are arranged on a circle controlled by two dimensions: valence (positive-negative) and arousal (weak-strong).
Consequently, each emotion category can be understood as a linear combination of these two dimensions, or as varying degrees of both valence and arousal.
While most emotional theories consider affective states from the point of view of psychology and psychiatry, machine learning systems usually takes one input among speech, vision, or physiological signals.
More precisely, a Speech Emotion Recognition (SER) system consider that emotion in speech is conveyed by both linguistic and acoustic modalities.
For example, Alva et al.~\cite{Alva2015} proved that arousal is better recognized from acoustic features and valence from linguistic features.
Considering these facts, we investigated the fusion of both acoustic and linguistic modalities in our work as many studies have proven its utility in comprehension related domain~\cite{Wollmer2013,Alam2014,Atrey2010,Liu2018}. 
In a previous work~\cite{Macary2020allosat}, we defined a new axis within the circumplex model that goes from satisfaction to frustration through a neutral state in the middle.
This axis has been proposed for the specific analysis of customer relationships in the context of call-center conversations.

SER systems are subject to different forms of variability which make commercial applications difficult to set up.
The first variability lies in the references used to train the models: Emotion perception is highly subjective, and several manual annotations are required to reach a kind of ``ground truth''.
This variability is usually measured with annotator agreements (kappa values or correlation coefficients).
Second, the reliability of the performances increases with the number of audio samples used to evaluate the models. In SER, this number is relatively small due to the high data collection and annotation cost, therefore, confidence intervals for the performances of the systems are highly required.
Finally, the third variability comes from the initialization of the parameters of the models, and possible shuffle of the data during the training stage.
In this paper, we intend to bring some insights to these three forms of variability with the investigation of individual annotations, the systematic addition of confidence intervals to the regression performances, and the evaluation of initialization impact on the performances.

The different experiments detailed in this article conclude that the linguistic modality is the major contributor for satisfaction recognition in call-center conversations. A deep analysis on the linguistic content of some conversations is carried on to investigate intrepretable factors able to explain the high contribution of the linguistic modality for this specific task. 

%
%
The main contributions of our study are the followings:
\begin{itemize}
    \item The use of pre-trained models for satisfaction recognition
    \item The fusion of acoustic and linguistic modalities
    \item The addition of protocols to evaluate performance variabilities (annotation, initialization, confidence intervals).
    \item The proposition of interpretable linguistic cues which explain the performances of our model
\end{itemize}

The paper is organized as follows: Section~\ref{sec:relatedWorks} presents the related works, followed by our motivations in Section~\ref{sec:motivation} and the global overview of our experimental protocols in Section~\ref{sec:overview}. Satisfaction recognition using either acoustic or linguistic modalities experiments and results detailed in Section~\ref{sec:features}, while the fusion experiments and results are described in Section~\ref{sec:both_modalities}. Section~\ref{sec:analysis} presents a complete analysis regarding annotation and linguistic content. The conclusion is drawn in the last Section.

\section{Related Works}
\label{sec:relatedWorks}

\subsection{Features for speech emotion recognition}
 Looking for speech cues that gives the best emotion recognition model has always been a ``holy grail''~\cite{Batliner2011,Eyben2016, Tahon2016}.
 However most studies agree that emotion mainly lies in prosody which is a combination of different factors such as intensity, intonation, rhythm and voice quality. These high level factors are usually estimated from low level descriptors: pitch, spectral features, MFCCs, energy, etc...
 Therefore, to analyze emotion in speech, researchers usually rely on various voice parameters set that are related to emotion~\cite{Ververidis2004,Vogt2005,Kwon2003} including fundamental frequency, speech rate, pauses, voice intensity, voice onset time, jitter (pitch perturbations), shimmer (loudness perturbations), voice breaks, pitch jumps, and measures of voice quality.
Para-linguistic sets used in Speech Emotion Recognition (SER) such as ComParE~\cite{Schuller2014}, and GeMAPS~\cite{Eyben2016} used in Interspeech Emotion Challenges~\cite{Schuller2013}, are designed to capture prosody.
Other features like spectral ones can also be extracted: among them, mel frequency cepstral coefficients (MFCCs) are clearly the most often used as they are robust to noisy signals, even if they have not been designed to retrieve prosodic information nor emotion as concluded in Tahon et al.~\cite{Tahon2016}.

For a while, SER has been dominated by the acoustic modality. However, emotions are not only conveyed by prosody but also by words. 
While automatic speech recognition systems (ASR) are more and more efficient, linguistic features can be extracted with high reliability. 
In the field of Sentiment Analysis (SA) where the goal is to find emotion in written text, different features were proposed such as POS-tagged (Part-Of-Speech-tagged) words~\cite{Vanaja2018,Dhar2018}, polarity dictionaries (SenticNet~\cite{Poria2013}, FAN~\cite{Monnier2014}) or features extracted with the GloVe representation~\cite{Meisheri2018} n-grams/bag-of-words~\cite{Chaffar2011}.
It should be noticed however, that spoken language differs from written text in the grammatical correctness, disfluences and non-verbal vocalizations such as laughter, breathing, and so on~\cite{Schuller2018}.


\subsection{Modality fusion}
Due to the small amounts of training data, SER has late moved to the neural paradigm. 
First studies have used RNN (Recurrent Neural Networks), especially with LSTMs (Long Short Term memory) to retrieve emotional categories~\cite{Lee2015} or continuous dimensions~\cite{Trigeorgis2016, Schmitt2018}. 
CNNs (Convolutional networks) have also been used to predict SEWA continuous dimensions~\cite{Schmitt2019} however LSTMs seems to better generalize on call-center data~\cite{Macary2020convorrec}. 

In order to take advantage of the linguistic content in SER, the fusion of both textual and audio information gains on popularity~\cite{Atmaja2020,Yoon2018,Sahu2019}. 
Three strategies are usually applied for multi-modal fusion: (a) at the feature level by concatenating the inputs of different modalities, (b) at the decision level with majority voting, or (c) at the model level by merging intermediate representations~\cite{Wollmer2013,Liu2018,Sebastian2019,Planet2012}. 
More precisely, the fused model (c) is done by concatenated outputs of two distinct networks corresponding to each modality to feed next layers~\cite{Cai2019}.
Many other modalities can be used in SER to better represent affective states. 
For example, audio representation, facial cues from video, textual information are used in the work of Chen et al.~\cite{Chen2017} and Poria et al.~\cite{Poria2017} while Wu et al.~\cite{Wu2011} focuses on semantics labels and audio features. Modality fusion always improve the performances obtained on speech only.

\subsection{Pre-trained features for NLP}
Expert features have the advantage to convey human understandable information but there are not the only way to represent data.
From other research domains such as SA, we assist to the rising of pre-trained self-supervised feature to represent the data, especially with word embeddings such as GloVe~\cite{Meisheri2018} or Word2Vec~\cite{Barhoumi2020}.
As there are trained on a massive amount of data, they tend to be able to efficiently represent data, without the need of human annotation.
Very recently, these pre-trained features spread in SER. 
Atmaja et al.~\cite{Atmaja2019} uses acoustic features consisting mostly in time and spectral domain features, and Word2Vec embedding for the linguistic part by performing a feature level fusion.
While Yenigalla~\cite{Yenigalla2018} et al. uses spectrogram and phoneme embeddings merged at the model level.

The self-supervised learning of speech or language representations has been proposed these last few years, for instance with the BERT system~\cite{Devlin2019}, used for textual representation.
Such representations, computed by neural models trained on huge amounts of unlabeled data, have shown their effectiveness on some tasks under certain conditions, for instance in ASR~\cite{Kahn2020,Liu2020}, or speech translation~\cite{Nguyen2020}. 
Recently Wav2Vec~\cite{Schneider2019}, Mockingjay~\cite{Liu2020} and Audio AlBERT~\cite{Chi2020} were introduced in ASR and speaker identification as one of the first pre-trained approaches to extract context dependent features from raw signals for ASR tasks but they have not been used for SER yet.
Very recently a BERT-like model for French has been developed~~\cite{Le2020}.
To the best of the authors' knowledge, such pre-trained features have not been yet used for SER.


\section{Motivation}\label{sec:motivation}

    The goal of our research is to continuously recognize satisfaction and frustration in real-life call-center conversations.
    To do so, we are using AlloSat~\cite{Macary2020allosat} French corpus to train speech emotion recognition network.

    
Moreover, several studies point out that emotion information can be detected not only in speech but also in facial traits, biological responses or linguistic and semantic information.
Traditionally, emotion recognition models use only the acoustic modality~\cite{Patel2013}, even if some works have shown that linguistic modality also convey important information~\cite{Devillers2006}. 
In our work, we investigate the use of the acoustic signal and its linguistic transcription, separately or jointly.
To compensate the lack of training data dedicated to the targeted task, we also explore the benefit of using models pre-trained on huge amount of data for both modalities such as Wav2Vec~\cite{Schneider2019}, Word2Vec~\cite{Mikolov2013} or BERT~\cite{martin2020}.

    
To design application for real industrial end users, one of the main concern is to be able to reproduce the results on multiple GPU clusters, thus to reduce all possible variabilities during the evaluation process.
In the scope of neural networks paradigm, weights initialization has always been pointed out as crucial as it impacts both the training time and the phenomena of being stuck in a local minima~\cite{Hendrycks2019}.

Therefore we will estimate how much the weight initialization affects the performances of the satisfaction recognition.
Because the Test set is, of course, not representative of all possible realizations, we decided to include an confidence interval to our scores.
This aims at given an idea of how much the performances could vary when evaluating on different conversations, considering that all non-deterministic sources are fixed for that matter.
In the field of continuous emotion recognition, the reference generally consists of the averaged value over all annotators. 
In our study, we tackle the problem of the subjectivity of the annotation by considering individual annotation instead of the averaged reference.

Our major conclusion is that models learnt on features extracted from the transcripts only are very accurate in the  prediction of satisfaction and frustration.
Therefore, this work analyses the linguistic content, and proposes relevant linguistic clues which are strongly related with the perception of the emotional state.

\section{Global overview}\label{sec:overview}

This section presents the speech material used to train and evaluate the models and the general architecture of the neural network used for SER.

\subsection{Speech emotional data : AlloSat corpus}


While past emotional speech corpora were annotated with discrete emotion categories~\cite{Devillers2010,Morrison2007}, the current trend is to move towards continuous annotations of affective dimensions.
Among the most popular corpora annotated continuously, we can cite SEMAINE~\cite{McKeown2012} composed of English interactions with virtual or human operators, or RECOLA~\cite{Ringeval2013} targeting French dyadic online conversations.
Both corpora are annotated at least according to arousal and valence dimensions.
The recent cross-cultural Emotion Database SEWA~\cite{SEWA} was presented for the 2018 Audio/Visual Emotion Challenge~\cite{AVEC2018} which aimed to retrieve arousal, valence and liking dimensions from semi-supervised dyadic conversations. 

In order to fit with our target task, we choose to carry on our experiments on AlloSat corpus~\cite{Macary2020allosat} composed of real-life call-center conversations, annotated along the satisfaction axis.
%
AlloSat was precisely built to continuously predict the evolution of the customer satisfaction on call-centers audio recordings of French speaking adult callers (i.e. customers). 
Various information are asked by the callers: contract information, global details on the company, or complains.

All conversations were recorded at 8kHz between July 2017 and November 2018 in call-centers located in French-speaking countries.
The agents are employees of various companies in different domains, mainly energy, travel agency, real estate agency and insurance.
The two telephone channels were recorded separately. Due to commercial constraints, we discarded the part of the receiver (i.e agent). 
Consequently, there is no overlap in the conversations.

AlloSat contains 303 conversations for a total duration of 37h 23' as summarized in Table~\ref{sumUpAlloSat}.
There is generally one single speaker per conversation even if some conversations can involve multiple speakers, for instance when the caller gives the telephone to someone else.
In order to preserve the speakers' privacy, all personal information were obfuscated with a jazzy sound letting the annotator knows that there was private information at this very moment. 
This anonymization process ensures to respect the General Data Protection Regulation (GDPR) recommendation. 
Because we removed the agent speech, there can be long moments of silence in the remaining caller speech. 
To minimize the annotator effort, we decided to replace these silences by 2 seconds of white noise, allowing the annotators to identify these moments of silence. 
In order to avoid collecting too many conversations with poor emotional content, we decided to apply three selection criterion based on prosodic and linguistic content.
\begin{enumerate}
    \item Speech duration: conversations longer than 30 seconds containing more than three speech turns;
    \item Intonation: standard deviation of the fundamental frequency ($F_0$) over 40~Hz. $F_0$ is extracted with YAPPT algorithm~\cite{Zahorian2008} which is adapted to telephone signals;
    \item Linguistic valence: the valence score computed on the transcriptions is below 4.98 (negative) or above 5.02 (positive). Word scores are given by FAN French dictionary~\cite{Monnier2014} and unknown words are at 5.00.
\end{enumerate}

Emotion annotation is known to be a highly subjective task. To compensate for the subjectivity of the annotation task, 
three annotators rated continuously the 303 conversations along the satisfaction axis.
This axis range from frustration to satisfaction with a neutral state in the middle and is sampled every 0.25 seconds. 
Individual annotations were averaged to get a gold reference, used in the prediction task. For more details about the coherence of the annotations, please refer to our previous work~\cite{Macary2020allosat}.
An automatic transcription were provided by Allo-Media for each conversation.

\begin{table}
\centering
\caption{\label{sumUpAlloSat} Summary of AlloSat characteristics}
\begin{tabular}{ll}
    \hline \textbf{Statistics} & \textbf{Value} \\ 
    \hline
     number of conversations	&303	\\
    \hline
    number of speakers      &308   \\ 
    number of women         &191 	\\
    number of men		    &117	\\
    \hline
    total duration      	&37h23m27s \\
    \hline
    min duration conversations		&32s		\\
    max duration conversations		&41m		\\
    mean duration conversations		&7m24s     \\
    \hline
\end{tabular}

\end{table}
    
The corpus has been divided into three subsets: The train set contains 201 conversations corresponding to about 25h of audio signal and 16h of speech; The development set is composed of 42 conversations; and the test set contains 60 conversations. 
Both Development and Test sets are composed of about 6h of audio signal and 3h of speech.
    
\subsection{SER neural network model}

\subsubsection{Baseline architecture}

We designed a regressive baseline neurol network to continuously predict the satisfaction along the conversation. 
To do so, a recurrent network, inspired from~\cite{Schmitt2019}, is used for the prediction task using bidirectionnal Long Short-Term Memory units (biLSTM). 

\begin{figure}
        \centering
         \includegraphics[width=.99\linewidth]{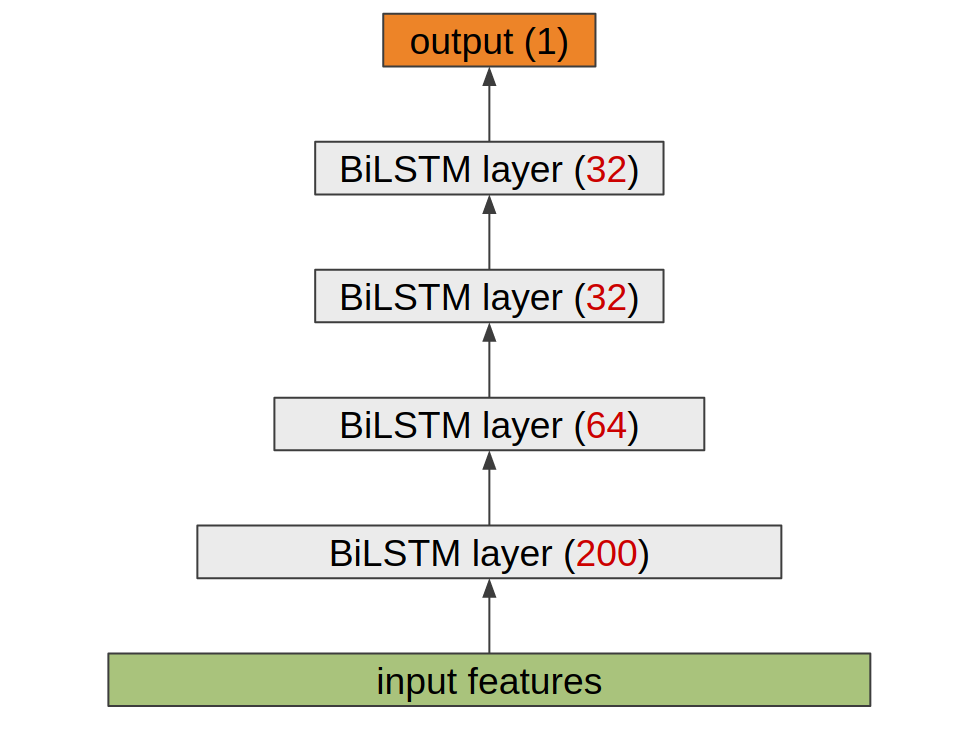} 
        \caption{Baseline network architecture. Number of neurons of each layer are written in red.} 
        \label{fig:models}
    \end{figure}

The sizes of the different layers have been optimized in our previous work, and the final architecture is composed of 4 biLSTM layers of respectively 200, 64, 32, 32 units with a $\tanh$ activation as shown on Figure~\ref{fig:models}.
A single output neuron is also used to predict the regression value each 250~ms at the emotional segment level.
Neither dropout nor batch normalisation is used in this approach.

The baseline network is fed with expert acoustic, respectively linguistic, feature sets of low dimension (40, respectively 48) described in the next section~\ref{sec:features}.
When moving to pre-train features, the input dimension explodes up to hundreds as they intend to represent huge amounts of speech data. 


A mean and variance normalization of the input features is done over the training data for all experiments.
    
\subsubsection{Loss and evaluation function}
The concordance correlation coefficient (CCC)~\cite{CCC} goes from 0 (chance level) to 1 (perfect) and is calculated according to eq.~\ref{eq:CCC_score}, where $x$ is the prediction and $y$ the reference. $\mu_x$ and $\mu_y$ are the means for the two variables and $\sigma_x$ and $\sigma_y$ their corresponding variances. $\rho$ is the correlation coefficient between the two variables $x$ and $y$.
        \begin{equation}
            CCC = \frac{2\rho\sigma_x\sigma_y}{\sigma_x^2 + \sigma_y^2 + (\mu_x - \mu_y)^2}
        \label{eq:CCC_score}
        \end{equation}
        
In previous experiments on the prediction of emotional dimensions~\cite{Trigeorgis2016,Schmitt2019}, the loss function to be minimized during the training phase is defined according to eq.~\ref{eq:loss}, where the CCC is computed over all concatenated conversations within a batch.
\begin{equation}\label{eq:loss}
    \mathcal{L}_c = 1 - CCC
\end{equation}

The CCC is also used as the evaluation metric on the Development and Test subsets. 
The score is computed at once on all the concatenated conversations of a given data subset, as described in AVEC challenges~\cite{AVEC2018}.

\subsubsection{Confidence interval for CCC score}
    
As mentioned previously, our work also intend to assess the robustness of the models from an industrial perspective.
More precisely, as the number of samples used to evaluate the models is relatively small, we need to estimate how reliable is the final CCC score with a confidence interval~\cite{Liao2000,McBride2005}.
The definition of the confidence interval for CCC is given in Appendix~\ref{ap:ccc}.
On AlloSat evaluations, the confidence interval widths for the CCC are between 0.006 (lower CCC) and 0.002 (high CCC). 
In the following experiments, a difference in performance will be judged as consistent if the two confidence intervals do not overlap.

\subsubsection{Hyper-parameters}
All networks are implemented under Pytorch framework~\cite{pytorch}. Preliminary experiments on the development set, helped to settle the baseline network architecture (number of biLSTM layers and number of neurons per layer) and the following hyper-parameters:
training is done on batches from 8 to 20 conversations 
using the Adam optimiser, depending on the size of the input embedding and memory constraints. All the conversations are kept without any padding.
The learning rate is optimized at 0.001 by empirical method, tested on a range from 0.001 to 0.02 by a 0.005 step. 
After preliminary experiments, we noticed that networks were not improving after the first 400 epochs, so the maximum number of epochs is set to 500.
For each training process, the final model is the one extracted from the epoch that gets the best score on the Development set. This final model is then evaluated on the Test set.


\subsubsection{Initialization}

the initialization of the model can have a huge impact on both the execution time and the accuracy of the resulting system. To handle with this hypothesis, 5 random initializations are tested on our best decision fusion system.
In additional experiments\footnote{The results are not presented here}, the final CCC score of one of the fusion approaches varies from $.873$ to $.911$ depending on the seed used for the initialization. It is a high variability which is considered to be relevant if we refer to the confidence interval, allowing us to conclude that the initialization is crucial.
In such a situation, if a new model is trained with same data and same architecture, there is a significant uncertainty on the final performances. This will not be investigated in the reste of the article.

\section{Acoustic and linguistic features}\label{sec:features}

This section describes features used in input of the network. While acoustic features are extracted directly from the speech signal, linguistic features are obtained from the transcription.
Baseline features consists of the traditional inputs used to represent the signal, \textit{i.e.} Mel Frequency Cepstral Coefficents (MFCCs), or textual information, \textit{i.e.} Word2Vec. Pre-trained features are indeed embeddings which are learnt on huge amount of data for an external task, here automatic speech recognition.


\subsection{Acoustic modality}

\subsubsection{Baseline features: MFCCs}

Two baseline acoustic sets are used as input of the network: MFCCs and the extended Geneva Minimalistic Acoustic Parameter Set (eGeMAPS). While eGeMAPS intend to precisely capture and represent prosody in speech, MFCCs are known to be robust to low quality audio signals such as telephone. In previous experiments, we have shown that MFCCs better achieve to predict satisfaction than eGeMAPS features~\cite{macary2021slt}, therefore only MFCCs are considered in the remainings.
%
In speech processing, the spectral content is considered as constant on small audio segments of around 30~ms. Our signal is sampled at 8~kHz, therefore MFCC 1-12 and their delta values are extracted on 30~ms frames each 10~ms with torchaudio toolkit\footnote{\url{https://pytorch.org/audio/stable/index.html}}.

Mean and standard deviation of each coefficient are computed over the emotional segment in order to get a 48 dimensional vector each 250~ms.
\subsubsection{Pre-trained features : Wav2Vec}
Self-supervised learning approaches have been designed in order to take benefit of huge amount of unlabelled data.
Wav2Vec (1.0)~\cite{Schneider2019} is a neural model trained through self-supervision to compute speech representations from raw audio.
This model is composed of two distinct convolutional neural networks. A first encoder network converts the audio signal into a new representation that is given to the second network, the ``context network'', which takes care of the context by aggregating multiple time step representations into a contextualized tensor that matches to a receptive field of about 210~ms. Both are then used to minimize a contrastive loss function. The resulting embedding is a 512-dimensional feature vector.
As the training of such model demands a lot of data and calculation power, we use the large pre-trained model provided by Schneider et al. in ~\cite{Schneider2019}, trained on Librispeech corpus~\cite{librispeech} consisting of 960 hours of English audio book samples at 16~kHz. Our features were extracted on an upsampled version of AlloSat\footnote{We used FFMpeg resampling function with \textit{sinc} interpolation function}.
In order to investigate the influence of the acoustic context on Wav2Vec representations, embeddings are extracted either on the current 250~ms emotional segment (without context) or on the whole conversation input (with context).
        
In the end, each emotional segment is represented by a 512-dimensional vector which consists of the averaged values of obtained embeddings over each segment of 250~ms.

\subsection{Linguistic modality}

    \subsubsection{Baseline features : Word2Vec}


Word2Vec embeddings have been extensively used for sentiment analysis or opinion mining from text~\cite{Barhoumi2020, Atmaja2019}, this motivated us to use such representation for the prediction of satisfaction.
In the following experiments, a Word2Vec model has been trained with the toolkit GENSIM~\cite{gensim}, using private data owned by Allo-Media composed of manual call transcriptions received by call centers, totaling over 500 hours of speech, with CBoW algorithm~\cite{Mikolov2013}. No stop list is used before extracting the embeddings.
In a first step, the output size embedding is fixed to 40 in order to have similar dimension with baseline MFCC features (\emph{i.e} 48). 
It is also motivated with empirical results showing that in the range between 20 and 60, the dimension 40 gave the best results.
We also did the experiment with a more standardized output size, fixed at 100.

\subsubsection{Pre-trained features : CamemBERT}
Inspired by RoBERTa~\cite{liu2019roberta} and BERT, CamemBERT~\cite{martin2020} is a multi-layer bidirectional Transformer.
CamemBERT is trained on the Masked Language Modeling (MLM) task which consists of replacing some tokens by either the token $<$MASK$>$ or a random token and asking the model to correct the tokens. The network uses a cross-entropy loss. The input consists of a mix of whole words and sub-words in order to take advantage of the context.
        
We use the ``camemBERT-base'' pre-trained model delivered by the authors and trained on the French part of OSCAR corpus~\cite{ortizsuarez2019} consisting of a set of monolingual corpora extracted from Common Crawl snapshot and totaling 138GB of raw text and 32.7B tokens after sub-word tokenization.
Text representations were extracted on Allosat by using this pre-trained model, and we summarized the results by averaging the continuous representations of sub-words occurring in the current emotional segment. 
In total, we use a 768-dimensional feature vector. In order to investigate the influence of the linguistic context on CamemBERT representations, embeddings are extracted either on the words pronounced during the current emotional segment (without context) or on the whole conversation input (with context).


\subsection{Results}

\begin{table}[hb!]
    \centering
    \caption{\label{tab:resultPretrain} Comparison of the audio and text modalities in terms of CCC computed on Development and Test sets on AlloSat. Shuffle is activated within batches. woc: without context; wc: with context. Relative difference between Dev and Test sets and relative improvement between baseline and pre-trained features, are given in \%.}
        \begin{tabular}{lll|ll l}
            \hline 
            && &\multicolumn{3}{c}{\textbf{Satisfaction}} \\
            \multicolumn{2}{l}{\textbf{Modality}} &\textbf{\# size} &\textbf{Dev} &\textbf{Test} & \textbf{Diff. (\%)}\\
            \hline
            \multicolumn{5}{l}{AUDIO} \\
            \hline
            &MFCC            &48 &.851 {\scriptsize [0.0]} &.651 {\scriptsize [0.0]} & -23.5\\
            &Wav2Vec woc &512 &.844 {\scriptsize [-0.8]} &\textbf{.806} {\scriptsize [+23.8]} & -4.5\\
            &Wav2Vec wc &512 &.823 {\scriptsize [-3.2]} & .656 {\scriptsize [+0.8]} & -20.3\\
            \hline
            \multicolumn{5}{l}{TEXT} \\
            \hline
            &Word2Vec        & 40 & .885 {\scriptsize [0.0]} & .861 {\scriptsize [0.0]} & -0.1\\
            &Word2Vec        & 100 & .853 {\scriptsize [-3.5]} & .812 {\scriptsize [-5.7]} & -4.7\\
            &CamemBERT woc  & 768 & .916 {\scriptsize [+3.5]} & .817 {\scriptsize [-5.2]} & -10.8\\
            &CamemBERT wc  & 768 & \textbf{.917} {\scriptsize [+3.7]} & \textbf{.924} {\scriptsize [+7.3]} & -0.8 \\
            \hline
        \end{tabular}
    
    \end{table}

All results on acoustic and linguistic modalities are reported in Table~\ref{tab:resultPretrain}.
We confirm that pre-trained features are achieving awesome results in comparison to baseline features. 
Especially, the performance impressively increases on the Test set (+23.8\%) when using Wav2Vec pre-trained features extracted without context (CCC=.806) instead of MFCCs features (CCC=.651). 
The relative improvement on Test set (+7.3\%) obtained when using CamemBERT pre-trained features extracted with context instead of Word2Vec is not as spectacular as the one obtained on acoustics because Word2Vec (CCC=.861) features already reach good results in comparison to MFCCs (CCC=.651).
However this modality seems more robust as it improves for both Dev and Test sets.

To confirm the reliability of our results, we can notice that the performance obtained by our models trained on acoustic features computed by the English Wav2Vec1.0 model is consistent, and even better, to the one obtained on the same data and presented in a recent study~\cite{evain2021task} that used a Wav2Vec2.0 model to extract acoustic features to feed smaller neural models.

Deeper experiments on the number of features used to train Word2Vec representations confirm that the best performance on Dev and Test set are obtained with a size of 40. 
Increasing the number of features to 100 degrades the score on Dev (-3.5\%) and Test (-5.7\%) sets. 
%
Regarding to confidence intervals detailed in appendix~\ref{ap:scores}, we confirm that all mentioned improvements are significant. 

A lot of differences exist by nature between CamemBERT and Word2Vec: complexity of the neural architecture, context-dependent dynamic embeddings \textit{vs.} static embeddings, sub-words \textit{vs.} words, \ldots 
The computation of CamemBERT needs a lot of GPUs, data and time.
However, we do not have the means to train such a model on specialized data with call-center conversations.
Fortunately with the help of the pre-trained model kindly distributed by the authors and it is possible to get very good results on the targeted SER task without owning such amount of resources. 


       
As described in Table~\ref{tab:resultPretrain}, the different acoustic and linguistic representations of the speech signal have different sizes which can impact the training of the network.
We previously investigated the impact of this dimension gap on system performances~\cite{macary2021slt}, by comparing the network presented in Figure~\ref{fig:models} with another one designed to reduce the dimension of input features. 
This reduction was done by adding an optional dense layer after the inputs and before the first biLSTM layer in order to reduce the input size to 40, resp. 48, for linguistic, resp. acoustic, modalities. 
We concluded that both architectures achieved comparable results and the addition of a dense layer was not necessary and that the input size does not significantly affect the results.

To conclude from Table~\ref{tab:resultPretrain}, we confirm the relevance of using pre-trained features for satisfaction recognition.
Surprisingly, we also found that linguistic embeddings, are able to capture a lot of emotional information directly from the transcribed speech as it performs slightly better than the acoustic one, especially when using CamemBERT features extracted with context.
At this point, we should notice that pre-trained linguistic features are extracted from textual transcriptions, however Word2Vec and CamemBERT models are trained on speech signals. 
Therefore some acoustic information (mainly phonetics) is, in a sense, also included in these linguistic features.
However, we do not know at this stage how prosodic and para-linguistic information is captured by pre-trained linguistic features.

\section{Modality fusion}
\label{sec:both_modalities}
    
    \begin{figure*}[t]
    \centering
        \begin{subfigure}{.49\textwidth}
          \centering
          \includegraphics[width=.7\linewidth]{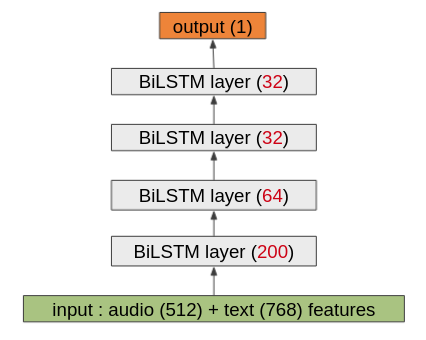}
          \caption{Feature fusion by concatenating input features.}\label{fig:fusion1}
        \end{subfigure}
        \hfill
        \begin{subfigure}{.49\textwidth}
          \centering
          \includegraphics[width=.88\linewidth]{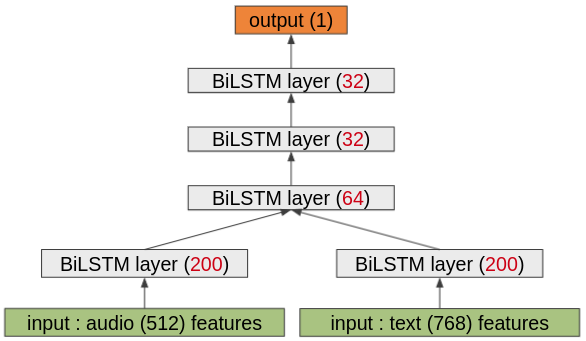}
          \caption{Model fusion by concatenating the first acoustic and linguistic layers.}\label{fig:fusion2}
        \end{subfigure}
        \hfill
        \begin{subfigure}{.49\textwidth}
          \centering
          \includegraphics[width=.88\linewidth]{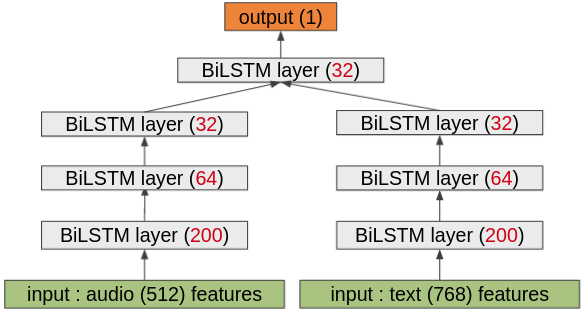}
          \caption{Model fusion by concatenating the last acoustic and linguistic layers.}\label{fig:fusion3}
        \end{subfigure}
        \hfill
        \begin{subfigure}{.49\textwidth}
          \centering
          \includegraphics[width=.88\linewidth]{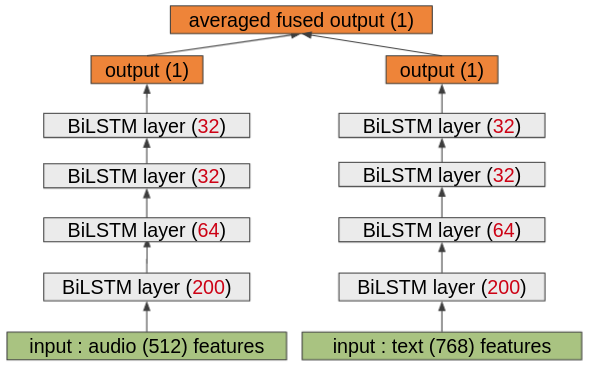}
          \caption{Decision fusion by averaging predictions from audio and text modalities.}\label{fig:fusion4}
        \end{subfigure}
    \caption{Description of the four used fusions.}
    \label{fig:fusion}
    \end{figure*}
            
As discussed in Section~\ref{sec:relatedWorks}, many studies confirm that emotion is conveyed by many modalities, especially acoustic and linguistic modalities as presented in Section~\ref{sec:relatedWorks}.
However, there is no consensus on the independence of acoustic and linguistic modalities, or there synchronicity with time.
To address this problem, we experiment three types of fusion : feature, model and decision fusion. 
In our case, the output value is return each 250~ms. 
Therefore acoustic and linguistic vectors must be aligned together with respect to time.

\subsection{Feature fusion}
Feature fusion methods enable a new representation of the speech signal which is the concatenation of individual modality features from the two modalities (Fig.~\ref{fig:fusion1}). 
A single model is then trained with a unique vector corresponding to a joint representation of the acoustic and linguistic features. The input size is therefore the sum of the two acoustic and linguistic feature sizes.
Good fusion performances at the feature level would probably mean that acoustic and linguistic modalities are synchronously used to perceive the satisfaction. 
        
\subsection{Model fusion}
We experiment two types of model fusion :
\begin{itemize}

\item Early fusion: Outputs of the first layers of acoustic and linguistic modalities are concatenated to feed the second layer (Fig.~\ref{fig:fusion2}).
\item Late fusion: Outputs of the third layers of acoustic and linguistic modalities are concatenated to feed the last biLSTM layer (Fig.~\ref{fig:fusion3}).
\end{itemize}

\subsection{Decision fusion}
To perform a decision fusion, two models are trained independently on each modality and the predicted numerical values are averaged to compute new predictions (Fig.~\ref{fig:fusion4}).  
In this configuration, it can be relevant to computed the global prediction (CCC$_G$) as the weighted average (Eq.~\ref{eq:decision_fusion}) of the individual acoustic CCC$_a$ and the linguistic CCC$_b$ scores, in order to give more importance to one of the two modalities.
\begin{equation}\label{eq:decision_fusion}
    \text{CCC}_G = w_a \cdot \text{CCC}_{a} + w_l \cdot \text{CCC}_{l}
\end{equation}

We optimize the weights of each modality from 0.1 to 0.9 with a step of 0.01.
The final configuration is the one which gives the better score on the development set.
Good fusion performances at the decision levels would probably mean that synchronicity is useless for the perception of satisfaction on a 250~ms frame.
    
        \subsection{Results}

The CCC scores obtained on Dev and Test sets with baseline (resp. pre-trained) features are summarized in Table~\ref{tab:resultFusionBaseline} (resp. Table~\ref{tab:resultFusion}). The relative differences between Test and Dev results are given in the last column to estimate the generalization power of the model.
Relative improvements are also included with the best single model as reference, \textit{i.e.} Word2Vec or CamemBERT.
Detailed scores with confidence interval can be found in appendix~\ref{ap:scores}.
        
\begin{table}[b!]
        \centering
        \caption{\label{tab:resultFusionBaseline} Comparison of four fusion approaches. CCC results on Dev and Test sets.
        Shuffle is activated within batches. 
        Relative differences between Dev and Test sets and relative improvements between baseline and pre-trained features, are given in \%.
        }
        \begin{tabular}{ll | ccc}
            \hline 
            &&\multicolumn{3}{c}{\textbf{Satisfaction}} \\
            &\textbf{Fusion level} &\textbf{Dev} &\textbf{Test} &\textbf{Diff. (\%)}\\ \hline
            \multicolumn{5}{l}{SINGLE BEST} \\ \hline
            &MFCC &.851~~~~~~~   &.651~~~~~~~ & -23.5\\
            &Word2Vec  & .883 {\scriptsize [0.0]} & \textbf{.881} {\scriptsize [0.0]} & -0.1 \\ \hline
            \multicolumn{5}{l}{FEATURE} \\ \hline
            
            &MFCC $\oplus$ Word2Vec     &.895 {\scriptsize [+1.4]}   &.833 {\scriptsize [-5.6]} & -6.9\\
            \hline
            \multicolumn{5}{l}{MODEL} \\ \hline
            &Early   &.904 {\scriptsize [+2.4]} &.807 {\scriptsize [-8.5]} & -10.7\\
            &Late    & \textbf{.917} {\scriptsize [+3.9]} &.815 {\scriptsize [-7.6]} & -11.1\\
            \hline
            \multicolumn{5}{l}{DECISION} \\ \hline
            & .66 Word2Vec + .34 MFCC    &.897 {\scriptsize [+1.6]} &.840 {\scriptsize [-4.8]} & -6.4\\
            \hline
        \end{tabular}
        
\end{table}

\begin{table}[t!]
        \centering
        \caption{\label{tab:resultFusion} Comparison of four fusion approaches. CCC results on Dev and Test sets.
        Shuffle is activated within batches.
        woc:w ithout context; wc: with context. 
        Relative differences between Dev and Test sets and relative improvements between baseline and pre-trained features, are given in \%.
        }
        \begin{tabular}{ll | ccc}
            \hline 
            &&\multicolumn{3}{c}{\textbf{Satisfaction}} \\
            &\textbf{Fusion level} &\textbf{Dev} &\textbf{Test} &\textbf{Diff. (\%)}\\ \hline
            \multicolumn{5}{l}{SINGLE BEST} \\ \hline
            &Wav2Vec woc &.844~~~~~~~   &.806~~~~~~~ & -4.5\\
            &CamemBERT wc  & .917 {\scriptsize [0.0]} & \textbf{.924} {\scriptsize [0.0]} & +0.8 \\ \hline
            \multicolumn{5}{l}{FEATURE} \\ \hline
            
            &Wav2Vec $\oplus$ CamemBERT     &.907 {\scriptsize [-1.1]}   &.884 {\scriptsize [-4.3]} & -2.5\\
            \hline
            \multicolumn{5}{l}{MODEL} \\ \hline
            &Early   &.924 {\scriptsize [+0.8]} &.897 {\scriptsize [-2.9]} & -2.9\\
            &Late    & \textbf{.945} {\scriptsize [+3.1]} &.893 {\scriptsize [-3.4]} & -5.5\\
            \hline
            \multicolumn{5}{l}{DECISION} \\ \hline
            & .72 CamemBERT + .28 Wav2Vec     &.932 {\scriptsize [+1.6]} &.920 {\scriptsize [-0.4]} & -1.3\\
            \hline
        \end{tabular}
        
        \end{table}

\begin{table*}[h!]
    \centering
    \caption{Fusion results for each annotator. Models are trained and evaluated on individual labels. CamemBERT are extracted with context while Wav2Vec are extracted without context. AVG: averaged over the three annotators. CV: coefficient of variation over the three annotators. Improvement corresponds to the absolute difference between CamemBERT and the decision fusion. Best fusion is chosen on Dev subset. Diff2: relative improvement between CamemBERT and best fusion.}
    \begin{tabular}{c | ll |cc| cc | cc | cc | cc}\hline
     &\multicolumn{2}{r|}{\textbf{Annotator}} & \multicolumn{2}{c|}{$a_1$} & \multicolumn{2}{c|}{$a_2$}  &\multicolumn{2}{c|}{$a_3$} &\multicolumn{2}{c|}{AVG} & \multicolumn{2}{c}{CV}\\
     Reference& \multicolumn{2}{l|}{\textbf{Fusion level}} & \textbf{Dev} &\textbf{Test} &\textbf{Dev}  &\textbf{Test} & \textbf{Dev} &\textbf{Test} & \textbf{Dev} &\textbf{Test} \\ \hline
     \multirow{9}{*}{\rotatebox{90}{\parbox{1.5cm}{Individual annotations}}} 
     &SINGLE    &Wav2Vec    & .834 & .734 & .731 & .785 & .841 & .597 & .802 & .705 & .077 & .138\\
     &          &CamemBERT  & .898 & .877 & .833 & .834 & .900 & .804 & .877 & .838 & .043 & .044\\
     &Diff1 (\%)&           & 7.7 & 19.5 & 14.0 & 6.2 & 7.0 & 34.7 & - & - & - & -\\
     \cline{2-13}
     &FEATURE   &           & .884 & .870 & .815 & .753 & .883 & \textbf{.834} & .861 & .819 & .046 & .073 \\
     \cline{2-13}
     &MODEL     & Early     & .883 & .870 & \textbf{.855} & \textbf{.865} & .888 & .826 & .875 & \textbf{.854} & .020 & .028\\
     &          & Late      & .911 & .875 & .814 & .837 & \textbf{.921} & .799 & .882 & .837 & .067 & .045\\
     \cline{2-13}
     &DECISION  &           & \textbf{.913} & \textbf{.882} & .840 & .849 & .916 & .793 & \textbf{.890} & .841 & .048 & .053 \\
    \cline{2-13}
     \hline\hline
     
     \multirow{9}{*}{\rotatebox{90}{\parbox{1.5cm}{Averaged annotations}}}
     &SINGLE    &Wav2Vec    & .862 & .736 & .774 & .731 & .779 & .710 & .805 & .726 & .061 & .019\\
     &          &CamemBERT  & .916 & .878 & .755 & .793 & .851 & .833 & .841 & .835 & .096 & .051\\
     &Diff1 (\%)&           & 6.3 & 19.3 & -2.5 & 8.5 & 9.2 & 17.3 & - & -\\
     \cline{2-13}
     &FEATURE   &           & .896 & .845 & .741 & .688 & .868 & .861 & .835 & .798 & .099 & .120 \\
     \cline{2-13}
     &MODEL     & Early     & .911 & .833 & \textbf{.809} & \textbf{.824} & \textbf{.879} & .856 & .866 & .838 & .060 & .020\\
     &          & Late      & .914 & \textbf{.899} & .763 & .784 & .844 & .841 & .840 & .841 & .090 & .068\\
     \cline{2-13}
     &DECISION  &           & \textbf{.938} & .882 & .795 & .778 & .868 & \textbf{.874} & \textbf{.867} & \textbf{.845} & .082 & .069 \\
    \cline{2-13}
     \hline
     
    \end{tabular}
    
    \label{tab:by_annotator}
\end{table*}

Table~\ref{tab:resultFusionBaseline} shows that whatever the fusion level, fusion performs better than Word2Vec only on the Dev set and lower on the Test set. The poor performances on the Test set can be explained by the very small CCC obtained with MFCCs (CCC=.651). The best improvement on the Dev set is obtained when using the late model fusion ($+3.9$\%), however this is the configuration that less generalizes on the Test set ($-11.1$\%). 

Table~\ref{tab:resultFusion} shows that the addition of the acoustic modality to CamemBERT embeddings does not improve performances on Dev set with feature fusion but with model or decision fusion. 
We confirm the fact that acoustic modality alone does not generalize well on the Test ($-4.5$\%) while CamemBERT does ($+0.8$\%).
The best improvement on Dev is obtained with a late model fusion ($+3.1$\%), however this is the configuration that less generalizes on the Test set ($-5.5$\%).
The decision fusion better generalizes on Test set ($-1.3$\%) than other fusion approaches and have the advantage of slighlty improving the Dev score ($+1.6$\%) while not much degrading on the Test ($-0.4$\%) in comparison to CamemBERT features only.

Unexpectedly, our results concludes that the linguistic modality (without the addition of acoustic features) best generalize to unseen data.
They also confirms the relevance of pre-trained features such as CamemBERT, to a lesser extent Wav2Vec, for satisfaction recognition in call-center converations.
While the model late fusion does not reach the best results in Test set, it significantly outperforms single linguistic modality on Dev, confirming the multi-modal aspects of emotion.
The advantage of this fusion method is that it requires less computing ressources to be trained.
Therefore, acoustic information is still useful but is less robust to unseen data.

\section{Analysis and discussion}\label{sec:analysis}
This section deeper analysis our results in order to better understand the importance of the linguistic modality. We investigates two axes: Annotator subjectivity and linguistic content.


\subsection{Influence of annotation subjectivity}

%
Our first analysis interrogates the subjectivity of the annotation task regarding acoustic and linguistic modalities. 
To do so, we modify the reference: Instead of training a single model on the averaged value over the three annotators, we train three different models per annotator, in which each reference is the single values for this annotator.
The predictions of these models are evaluated regarding individual annotations (top part of Table~\ref{tab:by_annotator}) or the ground truth defined as the average of the three individual annotations (bottom part of Table~\ref{tab:by_annotator}).
The AVG column gives the average performance over the three individual models.
The CV column gives the coefficient of variation (standard deviation over mean) over the three individual models.
Diff1 is the relative difference between linguistic and acoustic taken independently and gives an idea of the gain per annotator.


\textbf{Individual annotations: }
From the upper part of Table~\ref{tab:by_annotator}, we can notice that the coefficient of variation (CV) for single features, is higher with acoustic features than with linguistic features when the references are individual annotations, especially on the Test set.
More precisely, regarding annotator $a_3$, the performance of the acoustic modality severally drops on the Test set (CCC=.597).
Our hypothesis is that the variability in the acoustic space is highly diverse, and the same acoustic realization might be perceived with different satisfaction levels by the same annotator, what produces bad performances on the acoustic modality.
In the previous Section~\ref{sec:both_modalities}, we have shown that the fusion of the modalities improves performances on Dev but degrades on Test. This is not true when models are train and evaluated on individual annotations: fusion improves performances in most configurations and the best performance in average is reached with the model early fusion (CCC=.854 on Test set). 
The improvement on Test is highest with annotator $a_2$ ($+3.7$\% with model early fusion).
This can be explained by the very small difference between the performances obtained on independent modalities for this annotator ($+6.2$\%), maybe indicating that both modalities carry different information for this specific annotator.

From these results, we hypothesize that, at the annotator level, acoustic and linguistic modalities convey complementary emotional information, however, while the linguistic part is well shared among annotators, the perception of the acoustic part seems quite individual.
Of course additional experiments with cross-annotations are needed to confirm this hypothesis.






\textbf{Averaged annotations: }
Regarding individual models evaluated with averaged annotations (bottom part of Table~\ref{tab:by_annotator}), we notice that annotator $a_2$ has the lowest performances when using only linguistic features. 
The model built upon this annotator reaches the lowest performances using any type of fusion on Dev and Test sets.
Thus confirming the importance of high linguistic performances for the general evaluation.
This result can be explained by the fact that among the three annotators, we have shown that $a_2$ had the lowest intra-annotator agreement (see~\cite{Macary2020allosat}).
We also confirm the fact that the fusion helps to improve the performances per annotator in all cases.
The early model fusion has the advantage of having higher averaged performances than CamemBERT and of being the model less affected by individual annotations (CV =0.020 on Test set).

From these experiments, we conclude that while the fusion approaches degrade the global performances in comparison to CamemBERT only (see Table~\ref{tab:resultFusion}), it seems that they are more robust to the subjectivity of the annotation task.
We found that the early model fusion was the best compromise between performance and robustness.
%
Our insights also interrogates the evaluation process using the average values of the three annotators: averaged values have no perceptive reality, but individual values do.

\subsection{Linguistic analysis}

In the context of call-center conversations, the experiments described below conclude that the satisfaction-frustration axis is more supported by linguistic than acoustic content. 
Regarding to the circumflex model, this axis is very close to the valence axis, what could explain in some extend the importance of word for the detection of satisfaction.
%
In this section, we intend to provide elements that could explain the importance of linguistics to retrieve the satisfaction.
This analysis have been done on 13 conversations selected in order to cover different dynamics of the satisfaction dimension: Globally flat, occurrences of high frustration (ground truth $<$ 4) and occurrences of strongly decreasing satisfaction (frustration drops).
The analysis has been done using the automatic transcription, the reference satisfaction annotation and tags corresponding to \textit{high frustration} and \textit{frustration drop}.

Our hypothesis is that frustrated speech mainly correspond to the accentuation of the oral phenomena.
Consequently, we specifically investigated the following orality clues:
\begin{itemize}
    \item Amount of disfluencies,
    \item Hesitations, repairs, repetitions, babbling,
    \item Importance of self-breaks defined as ``the points where the utterance flow is broken''~\cite{pallaud2019disfluent},
    \item Usage of interrogations and negations,
    \item Semantic evidences of frustration or unhappiness,
    \item Amount of meaningfull segments \textit{vs.} semantically empty segments.
\end{itemize}

Based on these clues, the analysis concludes to different observations. 
There are semantic evidences of frustration in the conversations such as the usage of the negation (\textit{ça ne m'amuse pas}, \textit{c'est inadmissible}), strong markers (\textit{c'est gonflé}, \textit{putain de ...}) and weak markers (\textit{quand même}, \textit{franchement}).
It seems also that the amount of meaningful segments, self-breaks and disfluencies, are generally correlated with high frustration or satisfaction drops. The syntactic structure of interrogative utterances seems also correlated with frustration.

\begin{table*}[ht]
    \centering
    \caption{Extract (137 - 166~sec.) from a conversation about a certified letter. Disfluencies: \textit{italic}; Hesitations, repairs, babbling: \underline{underline}; Semantic evidences of frustration: \textbf{bold}; self-breaks: //}
    \begin{tabular}{p{0.49\textwidth}|p{0.49\textwidth}}
    \textbf{French} & \textbf{English translation} \\ \hline
    - \textit{voilà} et \underline{la deuxième lettre}  // c'est pareil \textit{mais bon} \underline{cette lettre}  // \underline{elle} est où maintenant… pas comprendre pourquoi on n'a pas retiré \underline{la lettre}... \underline{la deuxième lettre}  // c'est pareil \textit{mais} \underline{elle} venait d'où  // \underline{cette lettre}... c’était \underline{qui}  // \underline{qui} a envoyé \underline{cette lettre}... parce que c'est important  // on est une société  // nous… \underline{quand on sait pas qui c'est} // ... \underline{comment on peut savoir qui c'est} \textit{ouais mais} \underline{\textbf{ça va pas du tout}} \textit{hein} \underline{\textbf{ça va pas du tout}}  // \underline{ça}
    & 
    - \textit{there we are} and \underline{the second letter} // it is the same \textit{but} yes \underline{this letter} // where is \underline{it} now ... not understand why no one removed \underline{this letter} ... \underline{the second letter} // it is the same but where does \underline{it} come from // \underline{this letter} ... it is \underline{who} // \underline{who} sent \underline{this letter} ... because it is important // we are a society // we ... when we don't know who it is // ... how can we know who it is \textit{yeah but} \underline{\textbf{it's not ok}} \textit{eh} \underline{\textit{it's not ok}} // \underline{it}
    \\ \hline
    \end{tabular}
    \label{tab:ex_transcription}
\end{table*}

In a second step, we intend to go further in this analysis with the automatic extraction of orality clues. 
Of course, moving from manual to automated extraction implies to do some choices in the definition of the clues.
Trying to model the amount of meaningful segments, we extract POS tags using MACAON~\cite{nasr2011macaon} directly from automatic transcriptions and compute the number of verbs and nouns with respect to time.
To capture the other orality clues, we decided to extract automatically the seven features mentioned in Table~\ref{tab:ex_features}.

\begin{figure*}
    \centering
    \includegraphics[width=\textwidth]{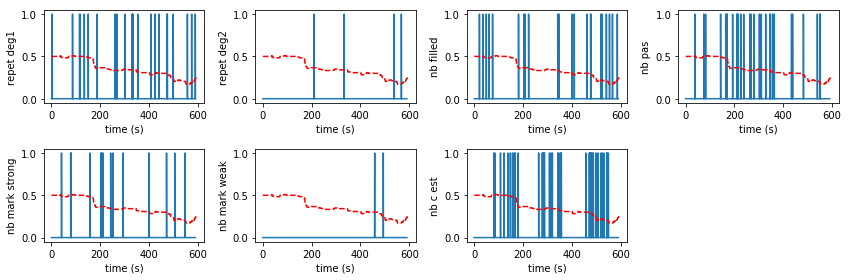}
    \caption{Dynamic analysis of frustration of a conversation about a certified letter. Number of occurrences of the seven linguistic features are plotted with respect to time. The gold satisfaction reference is represented with red dashed line.}
    \label{fig:ex_dynamic}
\end{figure*}

The idea is not to provide an exhaustive analysis on the whole dataset but to provide some explainable clues.
We focus here on the deep analysis of a single conversation about a certified letter.
All the occurrences of features summarized in Table~\ref{tab:ex_features} are synchronized in time together with the annotated satisfaction reference.
The number of verbs and nouns does not give relevant information and is not represented here.
The dynamic linguistic analysis of each conversation is shown on Fig.~\ref{fig:ex_dynamic}. This conversation has been annotated with a strong drop of satisfaction before 200~sec. The automatic transcription obtained just before this drop is given in Table~\ref{tab:ex_transcription}. Just before the drop, the occurrences of single words repetition and \textit{c'est} are important, whereas after the drop, the number of filled pauses and negation marker (\textit{pas}) increases. We also notice that a strong marker (\textit{réclamation}) happen just before the drop, probably meaning that this specific word induces the perception of noticeable frustration.

In the context of AlloSat speech data, emotional information seems to lies more in the words than in the prosodic and acoustic content. In such data, the expression of frustration is mainly related to the accentuation of the oral phenomena: semantic content and above all self-breaks, disfluencies, hesitations, repairs and repetitions.

\begin{table}[]
    \centering
    \caption{Seven features and their occurrences number used to model the orality clues supposed to be responsible for frustration in the conversations. The total number of utterances and words are included for the complete conversation about the certified letter.}
    \label{tab:ex_features}
    \begin{tabular}{l|c}
    \textbf{Features} & \textbf{\# occurrences} \\\hline
    single word repetitions (deg1)     & 26 \\
    bi-grams repetitions (deg2) & 4\\
    filled pauses (\textit{euh, bah, hein, eh, etc.}) & 22\\
    strong markers (\textit{important, inquiet, scandaleux, etc.}) & 14\\
    weak markers (\textit{quand même, franchement, etc.}) & 3\\
    negation marks (\textit{pas, ne, n'}) & 30\\
    \textit{c'est} & 44\\ \hline
    \# words in the conversation & 1050 \\
    \# utterances in the conversation & 152 \\ \hline
    \end{tabular}
    
\end{table}

\section{Conclusion}
This paper present the independent use of acoustic and linguistic pre-trained features and the fusion of these two modalities for the continuous recognition of satisfaction in call-center conversations.
We also present a further analysis on the influence of annotation subjectivity on the performances.
We also investigate possible linguistic clues able to explain the supremacy of linguistic features for this task.

Conducted on the AlloSat corpus, built for the recognition of satisfaction and frustration in real-life call-center conversations, we observe that Wav2Vec acoustic and CamemBERT linguistic pre-trained features, better represent satisfaction than baseline features such as MFCC and Word2Vec. On the Test set, the CCC score increases from  $0.651$ with 48 MFCC features to $0.806$ with Wav2Vec; and from $0.861$ with 40 Word2Vec to $0.904$ with CamemBERT.
In our experiments, we found that linguistic representations clearly outperform acoustic representations, thus questioning the need for acoustic in such task.
However, linguistic pre-trained features are extracted on automatic transcriptions directly obtained from the acoustic signals. 
So we definitely need acoustic and we do not know at this stage how prosodic and para-linguistic information is captured by these pre-trained features.

Our results clearly affirm the advantage of using CamemBERT representations, however the benefit of the fusion of acoustic and linguistic modalities is not as obvious.
With models learnt on individual annotations, we found that fusion approaches are more robust to the differences in annotations.
The early model fusion has the advantage of slightly degrading performances in comparison to model trained on CamemBERT features only, and being more robust to the subjectivity of the annotation task.

This article also investigates the robustness of the proposed approach towards industrial applications.
We pointed out the fact that the initialization process induces a large variability in the performances of the network. Further investigations are needed to cope with this issue.
We demonstrate that the use of fused models improves the robustness of the models regarding annotation subjectivity.
Finally a deep linguistic analysis allows us to propose relevant linguistic clues (negation and semantic markers, repetitions, filled pauses, etc.) that somewhat explains why the linguistic content is so important for this task.
We conclude that para-linguistic information is mainly included in words and their syntax.

In a future work, we intend to develop some approaches in order to cope with the initialization issue in order to provide reproducible experiments.
Additional experiments with cross-annotations approaches are needed in order to investigate the differences in perception due to the linguistic and acoustic modalities.
This work raises the question of the place of acoustic cues, especially prosody features. To pursue our investigation, we aim at applying the presented protocol on additional speech data, for instance broadcast news, political debates, etc.



\begin{table*}[h!]

\centering
\caption{\label{tab:resultConfidence} Confidence intervals}
        \begin{tabular}{ll | cc | cc}
            \hline 
            &&\multicolumn{4}{c}{\textbf{Satisfaction}} \\
            \multicolumn{2}{l|}{\textbf{Fusion level}} &\multicolumn{2}{c|}{\textbf{Dev}} &\multicolumn{2}{c}{\textbf{Test}} \\ \hline
            \multicolumn{5}{l}{SINGLE AUDIO} \\ \hline
            &MFCC         & .8507 & [.8491; .8523] & .6506 & [.6477; .6536]\\
            &Wav2Vec woc  & .8437 & [.8420; .8453] & .8055 & [.8036; .8073] \\
            &Wav2Vec wc   & .8234 & [.8215; .8252] & .6559 & [.6529; .6589]\\ \hline
            \multicolumn{5}{l}{SINGLE TEXT} \\ \hline
            &Word2Vec - 40  & .8848 & [.8836; .8860] & .8613 & [.8598; .8627] \\
            &Word2Vec - 100 & .8526 & [.8510; .8541] & .8124 & [.8105; .8143] \\
            &CamemBERT woc  & .9159 & [.9150; .9168] & .8166 & [.8148; .8185] \\
            &CamemBERT wc   & .9171 & [.9162; .9180] & .9239 & [.9231; .9248] \\ \hline
            \multicolumn{5}{l}{FEATURE FUSION} \\ \hline
            &MFCC $\oplus$ Word2Vec.- 40        &.8952 & [.8941; .8964] & .8331 & [.8315; .8348]\\
            &Wav2Vec woc $\oplus$ CamemBERT wc  &.9066 & [.9056; .9076] & .8840 & [.8828; .8851]\\
            \hline
            \multicolumn{5}{l}{MODEL FUSION} \\ \hline
            &MFCC $\oplus$ Word2Vec - 40 (early)   & .9039 & [.9028; .9049] & .8067 & [.8047; .8086] \\
            &MFCC $\oplus$ Word2Vec - 40 (late)    & .9166 & [.9157; .9175] & .8149 & [.8131; .8168] \\
            &Wav2Vec woc $\oplus$ CamemBERT wc (early)   & .8926 & [.8914; .8937] & .8937 & [.8926; .8947]\\
            &Wav2Vec woc $\oplus$ CamemBERT wc (late)    & .9450 & [.9444; .9456] & .8927 & [.8915; .8938] \\
            \hline
            \multicolumn{5}{l}{DECISION MODEL} \\ \hline
            & .66 Word2Vec-40 + .34 MFCC     & .9149 & [.9139; .9158] & .8351 & [.8334; .8367]\\
            & .72 CamemBERT + .28 Wav2Vec    & .9315 & [.9307; .9322] & .9202 & [.9194; 9210]\\
            \hline
        \end{tabular}

\end{table*}

\appendices
\section{Confidence interval for CCC}\label{ap:ccc}
Concordance correlation coefficient between two distributions $X$ and $Y$.
\begin{equation}
\hat{\rho_c} = \dfrac{2 \rho \sigma_x \sigma_y}{\sigma_x^2 + \sigma_y^2 + (\mu_x - \mu_y)^2} 
\end{equation}
Where:
\begin{itemize}
\item standard deviation: $\sigma_x = \dfrac{1}{N} \sum_{i} \left(x_i - \mu_x\right)^2$
\item covariance $\sigma_{xy} = \dfrac{1}{N} \sum_{i} \left(x_i - \mu_x\right)\left(y_i - \mu_y\right) $
\item correlation coefficient: $\rho = \dfrac{\sigma_{xy}}{\sigma_{x} \sigma_{y} }$
\end{itemize}

Applying the Fisher transformation is desirable to better meet the normal approximations. We call $\hat{Z}$ the estimator of the CCC~\cite{McBride2005}
\begin{equation}
\hat{Z} = \tanh^{-1}(\hat{\rho_c}) =  \dfrac{1}{2} \ln \left( \dfrac{1+ \hat{\rho_c} }{1 -\hat{\rho_c}}  \right)
\end{equation}

And the standard deviation of this estimate is:
\begin{equation}
\sigma_{\hat{Z} }^2 = \dfrac{\dfrac{(1-\rho^2) \hat{\rho_c} ^2}{(1-\hat{\rho_c}^2)\rho^2 } +  \dfrac{2\hat{\rho_c} ^3(1-\hat{\rho_c} )u^2}{\rho(1-\hat{\rho_c} ^2)^2} - \dfrac{\hat{\rho_c}^4 u^4}{2 \rho^2 (1-\hat{\rho_c}^2 )^2}}{N-2} 
\end{equation}
With the location shift relative to the scale parameter: $u = \dfrac{\mu_x - \mu_y}{\sigma_x \sigma_y}$.

Finally the confidence interval at 95\% for the CCC is:
\begin{equation}
    [\tanh (\hat{Z} - 1.64 \sigma_{\hat{Z}}); \tanh(\hat{Z} + 1.64 \sigma_{\hat{Z}})]
\end{equation}

\section{Full scores with confidence interval}
\label{ap:scores}

Table~\ref{tab:resultConfidence} summarizes the complete fusion CCC results with their confidence intervals. woc: without context, wc: with context.

\section*{Acknowledgments}

The authors thanks AlloMedia company for its technical support  in the collection of the AlloSat data, including its transcription.
The authors would like to thank Nathan Guilhot, internship at LIA in 2021, who helped towards the reproducibility of the results.

\ifCLASSOPTIONcaptionsoff
  \newpage
\fi

\bibliographystyle{IEEEtran}
\bibliography{bib}

\begin{IEEEbiography}[{\includegraphics[width=1in,height=1.25in,clip,keepaspectratio]{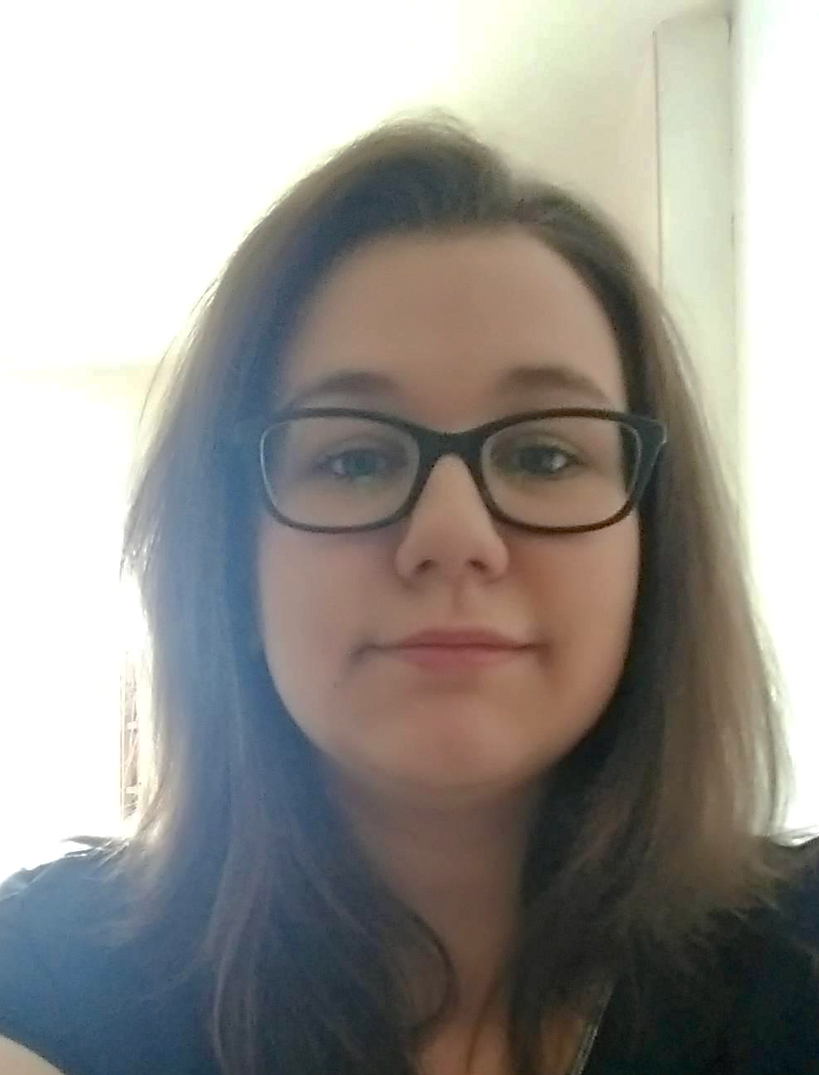}}]{Manon Macary}
received the master’s degree in Le Mans University (LIUM), where she is currently pursuing the Ph.D degree. She is also an employee of the Allo-Media company where she is active in the industrialization of her works. Her current research interests include emotion recognition from speech and spoken language understanding.
\end{IEEEbiography}


\begin{IEEEbiography}[{\includegraphics[width=1in,height=1.25in,clip,keepaspectratio]{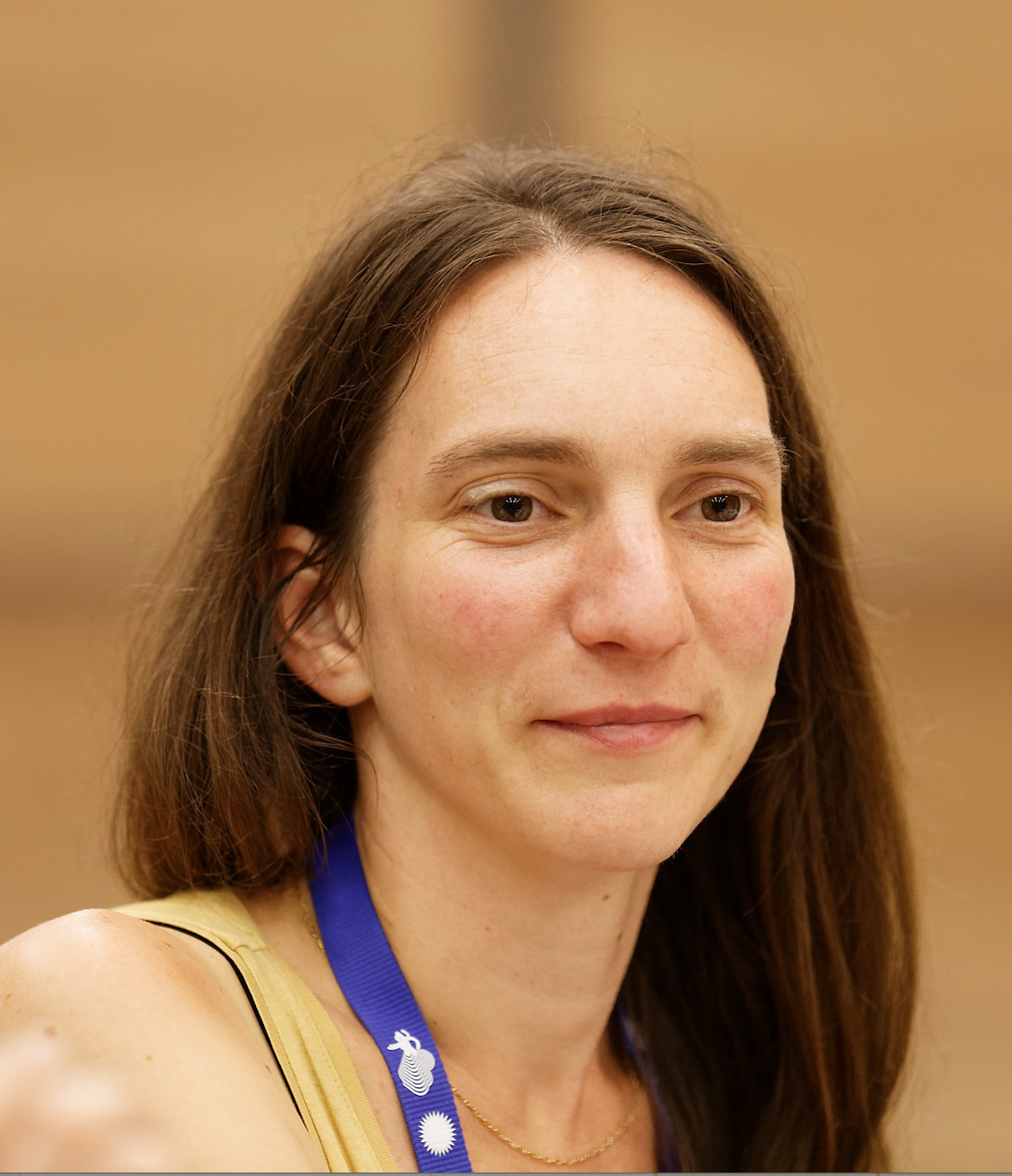}}]{Marie Tahon}
is currently Associate Professor at Le Mans University and conducts her research at LIUM (France). She graduated in engineering from the Ecole Centrale de Lyon (France) in 2007 and received the M.S. degree in acoustics from the Ecole Centrale de Lyon, in 2007. She received the Ph.D. degree in computer science from the University of Paris-Sud (Orsay, France) in 2012. She has been with the LIMSI-CNRS (Orsay, France) and with the IRISA (Lannion, France). Her research interests concern automatic speech processing for expressive speech: recognition and synthesis.
\end{IEEEbiography}


\begin{IEEEbiography}[{\includegraphics[height=1.25in,width=1in,clip,keepaspectratio]{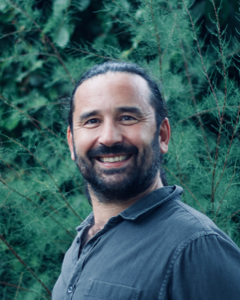}}]{Yannick Estève}
received the M.S. (1998) in computer science from the Aix-Marseilles University and the Ph.D. (2002) from Avignon University, France. 
He joined Le Mans Université (LIUM lab) in 2003 as an associate professor, and became a full professor in 2010. He moved to Avignon University in 2019 and is the head of the Computer Science Laboratory of Avignon (LIA) since 2020. He has authored and co-authored more than 150 journal and conference papers in speech and language processing.
\end{IEEEbiography}

\begin{IEEEbiography}[{\includegraphics[width=1in,height=1.25in,clip,keepaspectratio]{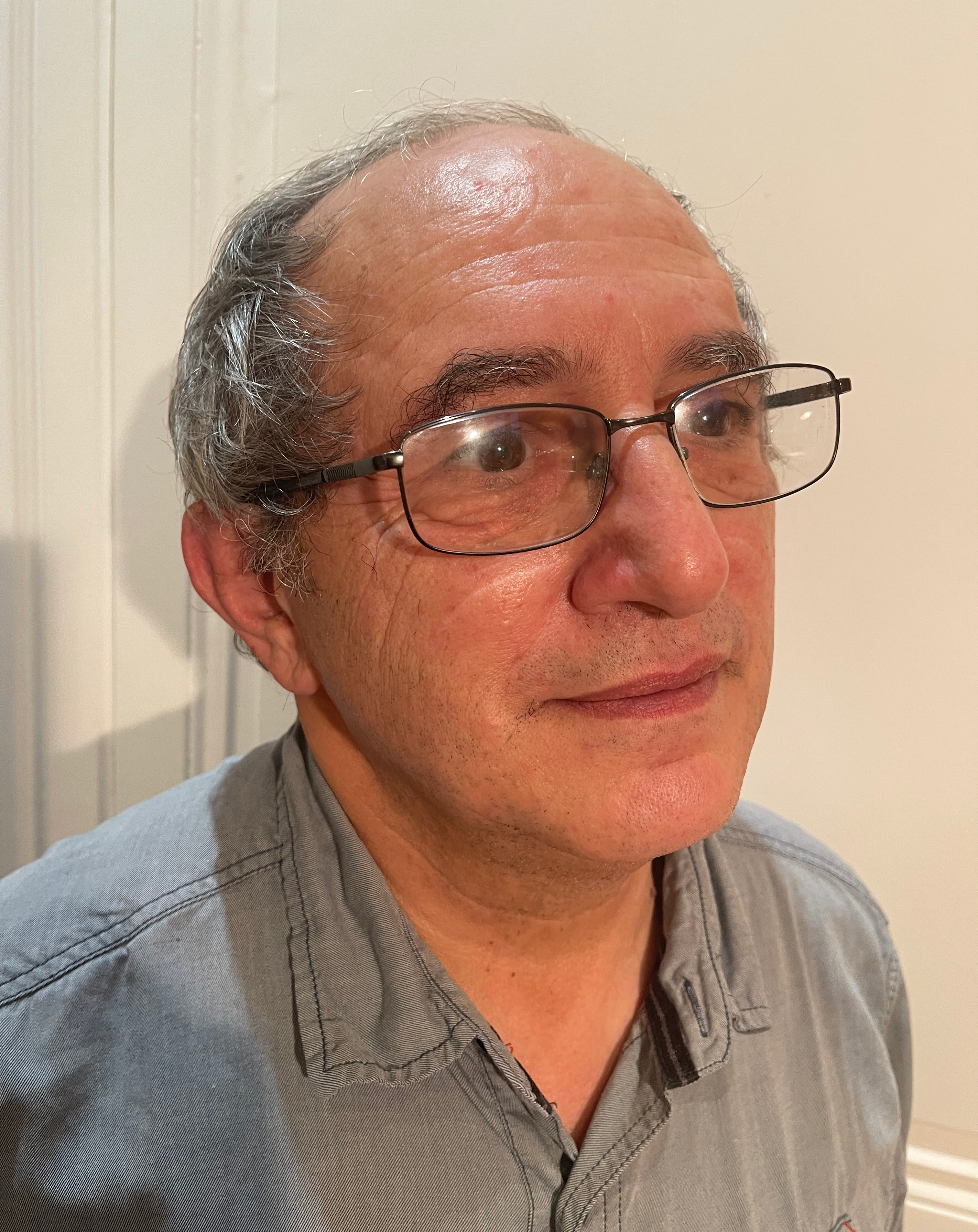}}]{Daniel Luzzati}
is currently Emeritus Professor at Le Mans University (LIUM). He performed two thesis, the first on spoken language (University of Paris 3), the second on human machine dialog (LIMSI-CNRS, Orsay). His research deals with these two areas, that affective computing both involves.
\end{IEEEbiography}

\vfill



\end{document}